\documentclass[12pt,showpacs,preprintnumbers,superscriptaddress,amsmath,amssymb,nofootinbib]{revtex4}
\usepackage{graphicx}
\usepackage{dcolumn}
\usepackage{bm}
\usepackage{amssymb}
\usepackage{amsmath}
\usepackage{epsfig}    
\usepackage{color}
\usepackage{hhline}

\def\be{\begin{equation}}
\def\ee{\end{equation}}
\newcommand{\bea}{\begin{eqnarray}}
\newcommand{\eea}{\end{eqnarray}}
\newcommand{\nn}{\nonumber}

\numberwithin{equation}{section}

\begin{document}

{\begin{flushright}{KIAS-P18072}
\end{flushright}}

\title{A predictive model of radiative neutrino mass with gauged $U(1)_{B-2L_{\ell_2}-L_{\ell_1}}$ symmetry }
%

\author{Takaaki Nomura}
\email{nomura@kias.re.kr}
\affiliation{School of Physics, KIAS, Seoul 02455, Republic of Korea}

\author{Hiroshi Okada}
\email{macokada.hiroshi@apctp.org}
\affiliation{Asia Pacific Center for Theoretical Physics, Pohang, Geyoengbuk 790-784, Republic of Korea}


\date{\today}

\begin{abstract}
We propose a predictively radiative neutrino mass model with gauged  $U(1)_{B-2L_{\ell_2}-L_{\ell_1}}$ symmetry in which we successfully realize two-zero textures of neutrino mass matrix depending on charge assignment for leptons and  the singlet scalar fields.  Then some correlations of neutrino parameters are shown where we take mass structure given by $(\ell_2, \ell_1) = (\mu, \tau)$ or $(\mu, e)$ as a representative case.  Furthermore we discuss phenomenology of the model such as lepton flavor violations, collider physics, and dark matter candidate.
\end{abstract}
\maketitle
\newpage

\section{Introduction}
It is known that {\it "radiative seesaw model"} is one of the elegant manners not only to explain small neutrino masses and dark matter candidate (DM) at the same time but also to correlate the neutrinos with DM, accommodating a lot of phenomenologies~\cite{Ma:2006km}.

On the other hand, recent experiments on neutrinos provide stringent constraints and the neutrino mass texture is tightly restricted.
For example, if one considers two zero neutrino mass textures with rank-three, only seven textures can survive~\cite{Fritzsch:2011qv, Alcaide:2018vni}.
Along this thought of idea, various unique textures can be realized by (Non-)Abelian symmetries in several frameworks of seesaw mechanisms such as type-II~\cite{Fritzsch:2011qv}, canonical seesaw~\cite{Kageyama:2002zw, Lamprea:2016egz, Borah:2017azf, Araki:2012ip, Asai:2017ryy}, inverse seesaw~\cite{Nomura:2018mwr, Dev:2017fdz}, linear seesaw~\cite{Sinha:2015ooa}, and so on. 
\footnote{Recently, analysis of the minimal textures(one zero textures with rank-two) has been done by~\cite{Barreiros:2018ndn}, and found  that only four textures are allowed all of which are favored of being inverted hierarchy.
Their realizations by symmetries have been done by ref.~\cite{Kobayashi:2018zpq} in basis of canonical seesaw.}

Combining a radiative seesaw model and a predictive neutrino mass texture is a natural subsequent task and several ideas are already realized in refs.~\cite{Baek:2015mna,Ko:2017quv, Lee:2017ekw, Nomura:2017ohi, Nomura:2018vfz}. 
 One can realize some specific neutrino mass textures by introducing lepton flavor dependent $U(1)$ gauge symmetry such as $U(1)_{L_\mu - L_\tau}$.
This possibility is interesting since a predictive neutrino mass texture can be related to phenomenology associated with $Z'$ boson. 

In this paper, we apply a flavor dependent gauged $U(1)_{B-2L_{\ell_2}-L_{\ell_1}}$ symmetry to find the predictive two zero textures to a radiative seesaw model at one-loop level based on ref.~\cite{Ma:2006km}. 
We discuss all the possibilities that can satisfy the current neutrino oscillation data~\cite{Gonzalez-Garcia:2014bfa} by relevant choice of charge assignment in lepton sector.
Then we focus on one of the promising textures, and investigate its phenomenologies such as lepton flavor violations (LFVs), collider physics and DM candidate as well as neutrino predictions.

This paper is organized as follows.
In Sec.~II, we explain our model, and then formulate each of sector such as Higgs sector, neutral fermions, active neutrinos,  LFVs,
and show all of the patterns that can satisfy the current neutrino oscillation data.
 In Sec.~III, we show numerical analyses of neutrino sector, collider physics, and DM candidate.
Finally we conclude in Sec.~IV.

 \section{Model: particle properties and phenomenologies}
\begin{center} 
\begin{table}
\begin{tabular}
{|c||c|c|c|c|c|c||c|c|c|}\hline\hline  
 & \multicolumn{6}{c||}{SM Leptons}   & \multicolumn{3}{c|}{Neutral fermions}\\\hline
Fermions  ~&~ $L_{L_{\ell_2}}$  ~&~ $L_{L_{\ell_1}}$ ~&~ $L_{L_{\ell_0}}$~ &~ $e_{R_{\ell_2} }$ ~&~ $e_{R_{\ell_1}}$&~ $e_{R_{\ell_0}}$~ &~ $N_{R_{\ell_2}}$&~ $N_{R_{\ell_1}}$ ~&~ $N_{R_{\ell_0}}$
\\\hline 
 $SU(2)_L$  & $\bm{2}$    & $\bm{2}$   & $\bm{2}$   & $\bm{1}$   & $\bm{1}$    & $\bm{1}$    & $\bm{1}$  & $\bm{1}$ & $\bm{1}$  \\\hline 
$U(1)_Y$  & $-\frac{1}{2}$  & $-\frac{1}{2}$  & $-\frac{1}{2}$   &  $-1$  &  $-1$ &  $-1$ & $0$ & $0$ & $0$  \\\hline
 $U(1)_{B-2L_{\ell_2}-L_{\ell_1}}$ & $-2$   & $-1$ & $0$  & $-2$  & $-1$ & $0$ & $-2$ & $-1$   & $0$  \\\hline
$Z_2$  & $+$  & $+$  & $+$ & $+$& $+$ & $+$  & $-$  & $-$ & $-$  \\\hline
\end{tabular} 
\caption{Field contents of fermions
and their charge assignments under $SU(2)_L\times U(1)_Y\times U(1)_{B-2L_{\ell_2}-L_{\ell_1}}\times Z_2$ ($\ell_2 \neq \ell_1$), where $SU(3)_C$ singlet for all leptonic fermions and all the quarks are assigned by $1/3$ charges under the $U(1)_{B-2L_{\ell_2}-L_{\ell_1}}$ as usual way. }
\label{tab:1}
\end{table}
\end{center}

\begin{table}[t]
\centering {\fontsize{10}{12}
\begin{tabular}{|c||c|c|c||c|}\hline\hline
&\multicolumn{3}{c||}{VEV$\neq 0$} & \multicolumn{1}{c|}{Inert } \\\hline
  Bosons  &~ $H$  ~ &~ $\varphi_{1(2)}$~ &~ $\varphi_{2(3)}$ ~&~ $\eta$ ~ \\\hline
$SU(2)_L$ & $\bm{2}$  & $\bm{1}$ & $1$  & $\frac12$     \\\hline 
$U(1)_Y$ & $\frac12$  & $0$ & $0$ & $\frac12$     \\\hline
 $U(1)_{B-2L_{\ell_2}-L_{\ell_1}}$ & $0$  & $1(2)$ & $2(3)$ & $0$   \\\hline
$Z_2$ & $+$ & $+$ & $+$  & $-$ \\\hline
\end{tabular}%
} 
\caption{Field contents of bosons
and their charge assignments under $SU(2)_L\times U(1)_Y\times U(1)_{B-2L_{\ell_2}-L_{\ell_1}}\times Z_2$, where $SU(3)_C$ singlet for all bosons. }
\label{tab:2}
\end{table}

{In this section, we introduce our models based on lepton flavor dependent $U(1)$ gauge symmetry, $U(1)_{B-2L_{\ell_2}-L_{\ell_1}}$, and formulate neutrino mass matrix and lepton flavor violating decay branching ratio of charged leptons.}

In the fermion sector, we introduce three right-handed Majorana fermions $N_{{e,\mu,\tau}}$ with isospin singlets that are needed to cancel our gauge anomalies, and impose the same charges as the SM lepton sector under the flavor dependent gauge symmetry  $U(1)_{B-2L_{\ell_2}-L_{\ell_1}}$ as summarized in Table~\ref{tab:1}.
 Also a discrete $Z_2$ symmetry is imposed for these new fermions in order to forbid the tree level neutrino masses and stabilize our DM candidate.

In the scalar sector, we add an inert $SU(2)_L$ doublet scalar $\eta$, and two singlet scalars $\varphi_{1,2 (2,3)}$ to the SM-like Higgs $H$ as summarized in Table~\ref{tab:2}, where the lower indices of $\varphi$
represents the charges of $U(1)_{B-2L_{\ell_2}-L_{\ell_1}}$ and $Z_2$ odd parity is  imposed to assure inert feature of $\eta$.
Notice here $\varphi_{i}$ have the vacuum expectation values (VEVs) after spontaneous symmetry breaking, which are respectively symbolized by $v_{\varphi_i}/\sqrt2$. 
We can realize different structures of neutrino mass matrix by choosing combination of singlet scalars $\{\varphi_1, \varphi_2 \}$ or $\{\varphi_2, \varphi_3 \}$ as we discuss below. 

{\it Yukawa interactions} : 
Under these fields and symmetries, the renormalizable relevant Lagrangian for neutrino sector and the Higgs potential is given by 
\begin{align}
-{\cal L}_L &= y_{\ell_2} \bar L_{L_{\ell_2}} H e_{R_{\ell_2}} +y_{\ell_1} \bar L_{L_{\ell_1}} H e_{R_{\ell_1}}
+y_{\ell_0} \bar L_{L_{\ell_0}} H e_{R_{\ell_0}} 
+ f_{\ell_2} \bar L_{L_{\ell_2}} \tilde\eta N_{R_{\ell_2}} + f_{\ell_1} \bar L_{L_{\ell_1}} \tilde\eta N_{R_{\ell_1}}   \nn\\
&
+ f_{\ell_0} \bar L_{L_{\ell_0}} \tilde\eta N_{R_{\ell_0}} + M_{\ell_0 \ell_0} \bar N_{R_{\ell_0}}^c N_{R_{\ell_0}}  
+ y_{N_{1(3)}} \varphi_{1(3)} \bar N_{R_{\ell_0(\ell_2)}}^c N_{R_{\ell_1}} \nn \\ 
& + y_{N_{2}} \varphi_{2} \bar N_{R_{\ell_1}}^c N_{R_{\ell_1}} 
+ y_{N_{3}} 
\varphi_{2} \bar N_{R_{\ell_0}}^c N_{R_{\ell_2}} 
+{\rm h.c.}
, \nn\\
V = & \mu_H^2 |H|^2 +  \mu_\eta^2 |\eta|^2 +
\frac{\lambda_H}{4}|H|^4 +\frac{\lambda_\eta}{4}|\eta|^4 
+ {\lambda_{H\eta}} |H|^2|\eta|^2 + {\lambda'_{H\eta}} |H^\dag\eta|^2
+ \frac{\lambda''_{H\eta}}4 [(\eta^\dag H)^2 + {\rm h.c.}]\nn\\
&+  \sum_{i=1}^{2(3)} \left(\mu_{\varphi_i}|\varphi_i|^2+
 \frac{\lambda_{\varphi_i}}{4}|\varphi_i|^4
+{\lambda_{H\varphi_i}} |H|^2 |\varphi_i|^2
+  {\lambda_{\eta\varphi_i}} |\eta|^2 |\varphi_i|^2 \right) + \mu (\varphi_2 \varphi_1^* \varphi_1^* + c.c.),
\label{eq:lag-quark}
\end{align}
where $\ell,\alpha,\beta$ respectively run over $(e,\mu,\tau)$, and 
$\tilde \eta\equiv i\sigma_2 \eta^*$, $\sigma_2$ being the second Pauli matrix.
Notice here that both $y$ and $f$ are diagonal due to our gauge symmetry, which is one of the important consequences.

For $\varphi_{1,2}$, we parametrize the scalar fields as 
\begin{align}
&H =\left[
\begin{array}{c}
w^+\\
\frac{v_H + h + iz_H }{\sqrt2}
\end{array}\right],\ 
{\eta} =\left[
\begin{array}{c}
\eta^{+}\\
\frac{\eta_R+i\eta_I}{\sqrt2}
\end{array}\right]
,\ 
\varphi_{1,2}=\frac{v_{\varphi_{1,2}}+\varphi_{R_{1,2}}+i\varphi_{I_{1,2}}}{\sqrt2},
\label{component}
\end{align}
where $v_H\simeq 246$ GeV is VEV of the SM-like Higgs, and $w^\pm$, $z$, and either of $\varphi_{I_{1,2}}$ are respectively GB 
which are absorbed by the longitudinal component of $W$, $Z$, and $Z'(\equiv Z_{B-2L_\ell-L_{\ell'}})$ boson.
Then we have the three by three CP-even  mass matrix squared $M^2_{R}$ in the basis of $[h, \varphi_{R_1},\varphi_{R_2}]^T$.
This is then diagonalized by $O_{R}^T M^2_R O_{R}\equiv$Diag[$m_{h_1},m_{h_2},m_{h_3}$],
where we identify $h_1\equiv h_{SM}$.
The situation is the same in $\varphi_{2,3}$ case.
The inert mass eigenstates are respectively written as
\begin{align}
&m_R^2(\equiv m^2_{\eta_R})=\mu_\eta^2 +\frac12(\lambda_{H\varphi_{1(3)}}v^2_{\varphi_{1(3)}}+\lambda_{H\varphi_2}v^2_{\varphi_2})
+\frac12(\lambda_{H\eta} +\lambda'_{H\eta}+\lambda''_{H\eta}/2)v_H^2 ,\\
&m_I^2(\equiv m^2_{\eta_I})=\mu_\eta^2 +\frac12(\lambda_{H\varphi_{1(3)}}v^2_{\varphi_{1(3)}}+\lambda_{H\varphi_2}v^2_{\varphi_2})
+\frac12(\lambda_{H\eta} +\lambda'_{H\eta}-\lambda''_{H\eta}/2)v_H^2 ,\\
&m_\eta(\equiv m_{\eta^\pm})=
\mu_\eta^2 +\frac12(\lambda_{H\varphi_{1(3)}}v^2_{\varphi_{1(3)}}+\lambda_{H\varphi_2}v^2_{\varphi_2})
+\frac12\lambda_{H\eta} v_H^2,
\end{align}
where inert conditions should also be imposed in the theory~\cite{Barbieri:2006dq}.

{We have heavy neutral gauge boson $Z'$ after $U(1)_{B-2L_\ell-L_{\ell'}}$ gauge symmetry breaking.
The mass of $Z'$ is given by
\begin{equation}
m_{Z'} = g' \sqrt{1(4)v_{\varphi_{1(2)}}^2 + 4(9)v_{\varphi_{2(3)}}^2},
\label{eq:MZp1}
\end{equation}
where $g'$ is the gauge coupling of $U(1)_{B-2L_{\ell_2}-L_{\ell_1}}$.}

\begin{table}[tb]
\centering {\fontsize{10}{12}
\begin{tabular}{|c||c|c|}\hline\hline
  $U(1)_{B-2L_{\ell_2}-L_{\ell_1}}$  &\multicolumn{2}{c||}{Texture of $M_N$}  \\\hline
$(\ell_2,\ell_1)$  &~ ($\varphi_1,\varphi_2$)  ~ &~  ($\varphi_2,\varphi_3$)~   \\\hline\hline
($\mu,\tau$) & $-$
& $\left(\begin{array}{ccc}
\times &  \times &  0 \\
\times & 0 & \times \\
0 & \times  & \times \\ 
 \end{array}\right)$ 
\\\hline
($\tau,\mu$) & $-$
& $\left(\begin{array}{ccc}
\times &  0 &  \times \\
0 & \times & \times \\
\times & \times  & 0 \\ 
 \end{array}\right)$ 
\\\hline
($e,\mu$)  & $\left(\begin{array}{ccc}
0 &  0 &  \times \\
0 & \times & \times \\
\times & \times  & \times \\ 
 \end{array}\right)$
& $-$
\\\hline
($\mu,e$)  & $\left(\begin{array}{ccc}
\times &  0 &  \times \\
0 & 0 & \times \\
\times & \times  & \times \\ 
 \end{array}\right)$
& $\left(\begin{array}{ccc}
\times &  \times &  0 \\
\times & 0 & \times \\
0 & \times  & \times \\ 
 \end{array}\right)$ 
\\\hline
($e,\tau$)  & $\left(\begin{array}{ccc}
0 &  \times &  0 \\
\times & \times & \times \\
0 & \times  & \times \\ 
 \end{array}\right)$
& $-$
\\\hline
($\tau,e$)  & $\left(\begin{array}{ccc}
\times &  \times &  0 \\
\times & \times & \times \\
0 & \times  & 0 \\ 
 \end{array}\right)$
& $\left(\begin{array}{ccc}
\times &  0 &  \times \\
0 & \times & \times \\
\times & \times  & 0 \\ 
 \end{array}\right)$ 
\\\hline
\end{tabular}%
} 
\caption{Possible two-zero mass textures of the right-handed neutrino that can realize active neutrino mass structure allowed by the current neutrino oscillation data.}
\label{tab:MN-texture}
\end{table}

{\it Neutral fermion masses}:
First of all, let us consider the masses of  Majorana fermions $N_R$,
where the mass texture depends on the assignment of $\ell_{1,2}$ and $\varphi_{1,2 (2,3)}$ as summarized in Table~\ref{tab:MN-texture}.
Here let us consider a texture with $U(1)_{B-2L_e-L_{\mu}}$ and $\varphi_{1,2}$, as an example. 
After the spontaneous symmetry breaking, in this case, the mass matrix is found to be 
 \begin{align}
&M_N =
\left[\begin{array}{ccc}  0 & 0 & \frac{y_{N_3}v_{\varphi_2}}{\sqrt2} \\
0&  \frac{y_{N_2}v_{\varphi_2}}{\sqrt2} &  \frac{y_{N_1}v_{\varphi_1}}{\sqrt2} \\
\frac{y_{N_3}v_{\varphi_2}}{\sqrt2} &  \frac{y_{N_1}v_{\varphi_1}}{\sqrt2}  & M_{\tau\tau} \\ 
 \end{array}\right]\equiv 
 \left[\begin{array}{ccc}  0 & 0 & M_{13} \\
0& M_{22} &  M_{23} \\
M_{13} & M_{23} & M_{\tau\tau} \\ 
 \end{array}\right].
\end{align}
Then we diagonalize $M_N$ as $V_N M_N V_N^T\equiv D_\psi$, where $V_N$ is unitary matrix in general and $N_R\equiv V_N^T \psi_R$, $D_\psi$ and $\psi_R$ being mass values and eigenstates, respectively.
The mass matrix can be treated by the same way in the other charge assignment in Table~\ref{tab:MN-texture}.

{\it Active neutrino masses}:
Now  we consider the active neutrino mass matrix at one-loop level~\cite{Ma:2006km}. 
It can be formulated by
\begin{align}
&  (m_\nu)_{\alpha\beta} = \sum_{i=1-3}
 \frac{f_\alpha V^T_{N_{\ell i}} D_{\psi_i} V_{N_{i\beta}} f_\beta}{2(4\pi)^2}
\left(\frac{m^2_R}{D^2_{\psi_i} - m^2_R}\ln\frac{m^2_R}{D_{\psi_i}^2}-
\frac{m^2_I}{D^2_{\psi_i} - m^2_I}\ln\frac{m^2_I}{D_{\psi_i}^2} \right).
\end{align}

In the case where $D_\psi^2 << m_0^2(\equiv (m_R^2+m_I^2)/2)$, the above formula simplifies as~\cite{Baek:2015mna} 
\begin{align}
&  (m_\nu)_{\alpha\beta} \approx \sum_{i=1-3} \frac{\lambda''_{H\eta} v_H^2}{2(4\pi)^2m_0^2}
 f_\alpha V^T_{N_{\ell i}} D_{\psi_i} V_{N_{i\beta}} f_\beta =
  \sum_{i=e,\mu} \frac{\lambda''_{H\eta} v_H^2}{2(4\pi)^2m_0^2}
 f_\alpha M^*_{N_{\alpha\beta}} f_\beta, \label{eq:type2-neut}
\end{align}
where we have used $V^T_N D_\psi V_N= M_N^*$ in the last form.
One finds that it is reduced to be a type-II form of  neutrino mass matrix.

In the case where $D_\psi^2\approx m_0^2$, the above formula simplifies as~\cite{Asai:2017ryy}  
\begin{align}
&  (m_\nu)_{\alpha\beta} \approx \sum_{i=1-3} \frac{\lambda''_{H\eta} v_H^2}{4(4\pi)^2}
{ f_\alpha V^T_{N_{\ell i}} D_{\psi_i}^{-1} V_{N_{i\beta}} f_\beta}  
\approx \sum_{i=e,\mu} \frac{\lambda''_{H\eta} v_H^2}{4(4\pi)^2}
{ f_\alpha  M_{N_{\alpha\beta}}^{-1} f_\beta}  , \label{eq:can-neut}
\end{align}
where we have used $M_N^{-1}=V^T_N D^{-1}_\psi V_N$ with $D_\psi^*=D_\psi$ in the last form.
\footnote{In the case where $D_\psi^2 >> m_0^2$, the above formula simplifies as 
$(m_\nu)_{\alpha\beta}$$\approx$$\sum_{i=1-3} \frac{\lambda''_{H\eta} v_H^2}{2(4\pi)^2}
\frac{ f_\alpha V^T_{N_{\ell i}}V_{N_{i\beta}} f_\beta}{D_{\psi_i}}\left[\ln\frac{D_{\psi_i}^2}{m^2_0}-1\right]$,
whose form gives a general symmetric matrix.}
It transversely suggests that
\begin{align}
&  M_N\approx \frac{\lambda''_{H\eta} v_H^2}{4(4\pi)^2}
{ f  m_{\nu}^{-1} f} . \label{eq:can-MN}
\end{align}

It is clearly that each of Eq.~(\ref{eq:type2-neut}) and~(\ref{eq:can-neut}) directly reflects on the structure of Majorana mass matrix;
$m_\nu \propto M_N^*,\ m_\nu \propto M_N^{-1}$,  
predictive two-zero textures are found depending on the mass hierarchy between $D_\psi$ and $m_0$.
In Table \ref{tab:MN-texture}, we have summarized all the possible textures that can reproduce the current neutrino oscillation data~\cite{Fritzsch:2011qv}.
In our numerical analysis, we will use the global fit of the current neutrino oscillation data at 3$\sigma$ confidence level for NH and IH~\cite{Gonzalez-Garcia:2014bfa}:
\begin{align}
{\rm NH}:\ 
&s_{12}^2=[0.270\sim0.344],\quad s_{23}^2=[0.382\sim0.643],\quad s_{13}^2=[0.0186\sim0.0250],\nn\\
&\Delta m^2_{21}\equiv m_{\nu_2}^2- m_{\nu_1}^2 =[7.02\sim8.09]\times10^{-5}\ {\rm eV}^2,\nn\\
&\Delta m^2_{31}\equiv m_{\nu_3}^2- m_{\nu_1}^2 =[2.317\sim2.607]\times10^{-3}\ {\rm eV}^2,\\
{\rm IH}:\ 
&s_{12}^2=[0.270\sim0.344],\quad s_{23}^2=[0.389\sim0.644],\quad s_{13}^2=[0.0188\sim0.0251],\nn\\
&\Delta m^2_{21}\equiv m_{\nu_2}^2- m_{\nu_1}^2 =[7.02\sim8.09]\times10^{-5}\ {\rm eV}^2,\nn\\
&\Delta m^2_{23}\equiv m_{\nu_2}^2- m_{\nu_3}^2 =[2.307\sim2.590]\times10^{-3}\ {\rm eV}^2,
\end{align}
where $s_{12,13,23}$ are the short-hand notations of $\sin\theta_{12,13,23}$ for three neutrino mixing angles of $U$.
Here we impose the experimental neutrino mass difference squares, and show what kinds of predictions can be found later.

In case of type-II in Eq.~(\ref{eq:type2-neut}),
one finds the following relations \cite{Fritzsch:2011qv} because of  $m_\nu\propto M_N^*$:
\begin{align}
\frac{m_{\nu_1}}{m_{\nu_3}}e^{2i\rho}&=
\frac{U_{a3} U_{b3}U_{\alpha2}U_{\beta2}-U_{a2} U_{b2}U_{\alpha3}U_{\beta3}}
{U_{a2} U_{b2}U_{\alpha1}U_{\beta1} - U_{a1} U_{b1}U_{\alpha2}U_{\beta2}},\\
\frac{m_{\nu_1}}{m_{\nu_3}}e^{2i\sigma}&=
\frac{U_{a1} U_{b1}U_{\alpha3}U_{\beta3}-U_{a3} U_{b3}U_{\alpha1}U_{\beta1}}
{U_{a2} U_{b2}U_{\alpha1}U_{\beta1} - U_{a1} U_{b1}U_{\alpha2}U_{\beta2}},
\label{eq:cond-tp2}
\end{align}
where indices for PMNS matrix corresponds to $(m_\nu)_{ab} = (m_\nu)_{\alpha \beta} = 0$ with $(ab) \neq (\alpha \beta)$, and $\rho$ and $\sigma$ are Majorana phases defined by $P=Diag(e^{i\rho},e^{i\sigma},1)$ with $V\equiv UP$.\\
In case of type of the canonical seesaw in Eq.~(\ref{eq:can-MN}), on the other hand,
one also finds the following relations because of  $M_N \propto m_\nu^{-1}$:
\begin{align}
\frac{m_{\nu_1}}{m_{\nu_3}}e^{2i\rho}&=
\frac{U_{a2}^* U_{b2}^* U_{\alpha1}^* U_{\beta1}^* - U_{a1}^* U_{b1}^* U_{\alpha2}^* U_{\beta2}^*}
{U_{a3}^* U_{b3}^*U_{\alpha2}^*U_{\beta2}^* - U_{a2}^* U_{b2}^* U_{\alpha3}^* U_{\beta3}^*}
,\\
\frac{m_{\nu_1}}{m_{\nu_3}}e^{2i\sigma}&=
\frac{U_{a2}^* U_{b2}^* U_{\alpha1}^* U_{\beta1}^* - U_{a1}^* U_{b1}^* U_{\alpha2}^* U_{\beta2}^*}
{U_{a1}^* U_{b1}^* U_{\alpha3}^* U_{\beta3}^* - U_{a3}^* U_{b3}^* U_{\alpha1}^* U_{\beta1}^*}.\label{eq:cond-cs}
\end{align}

{\it Lepton flavor violations(LFVs)}:
Here, we have to always be taken into account in any radiative seesaw models.
The formula is given by 
\begin{align}
&{\rm BR}(\ell_b\to\ell_a\gamma)\approx\frac{48\pi^3\alpha_{em}C_{ba}}{(4\pi)^4 G_F^2}
\left| {f_a V^T_{N_{ai}} V^*_{N_{ib}}f^*_b}{} F_{II}(D_{\psi_i},m_\eta)\right|^2,
\\
& F_{II}(m_a,m_b)\approx \frac{2 m_a^6 + 3m_a^4m_b^2-6m_a^2m_v^4+m_b^6+12m_a^4m_b^2\ln\frac{m_b}{m_a}}{12(m_a^2-m_b^2)^4},
\end{align}
where $G_F$ is Fermi constant, $\alpha_{em}$ is fine structure constant, and $C_{\mu e}\approx1$, $C_{\tau e}\approx0.1784$,
$C_{\tau\mu}\approx0.1736$. 
The current experimental bounds are given by 
\begin{align}
{\rm BR}(\mu\to e\gamma)\lesssim4.2\times10^{-13},\quad {\rm BR}(\tau\to e\gamma)\lesssim4.4\times10^{-8},\quad
{\rm BR}(\tau\to \mu\gamma)\lesssim3.3\times10^{-8}.
\end{align}
These constraints are imposed in considering neutrino oscillations.

\section{Numerical analysis and phenomenology of the model}

\subsection{Numerical analysis for neutrino mass matrix}\label{na}
Here we focus on the type of $B_1$; 2-2 and 1-3 components are zero~\cite{Fritzsch:2011qv} in Table \ref{tab:MN-texture}  obtained from charge assignment $(\ell_2, \ell_1) = (\mu, \tau)$ or $(\mu, e)$, and analyze some predictions of neutrino oscillation data in the case of type-II~in Eq.~(\ref{eq:cond-tp2}) and canonical seesaw~in Eq.~(\ref{eq:cond-cs}).

\begin{figure}[h!]
\begin{center}
\includegraphics[width=80mm]{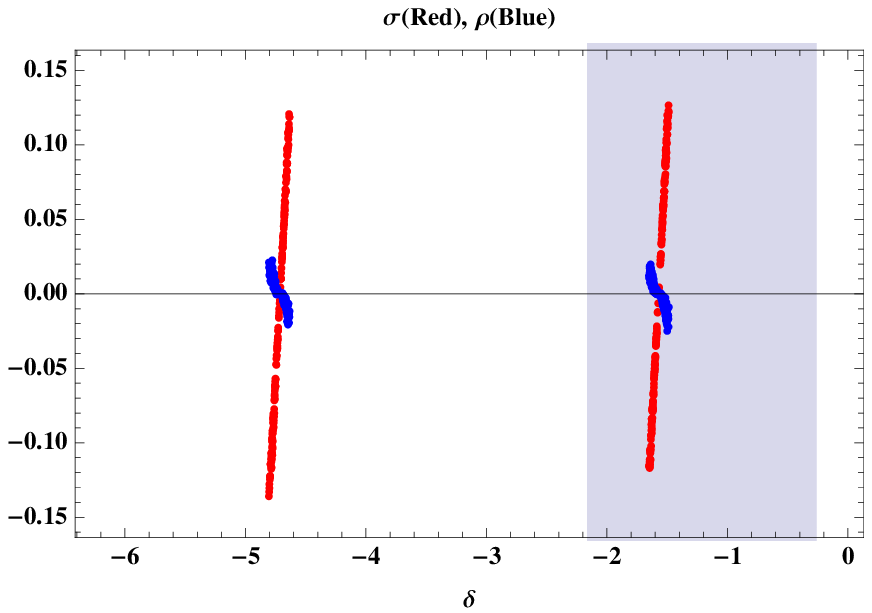} 
\includegraphics[width=80mm]{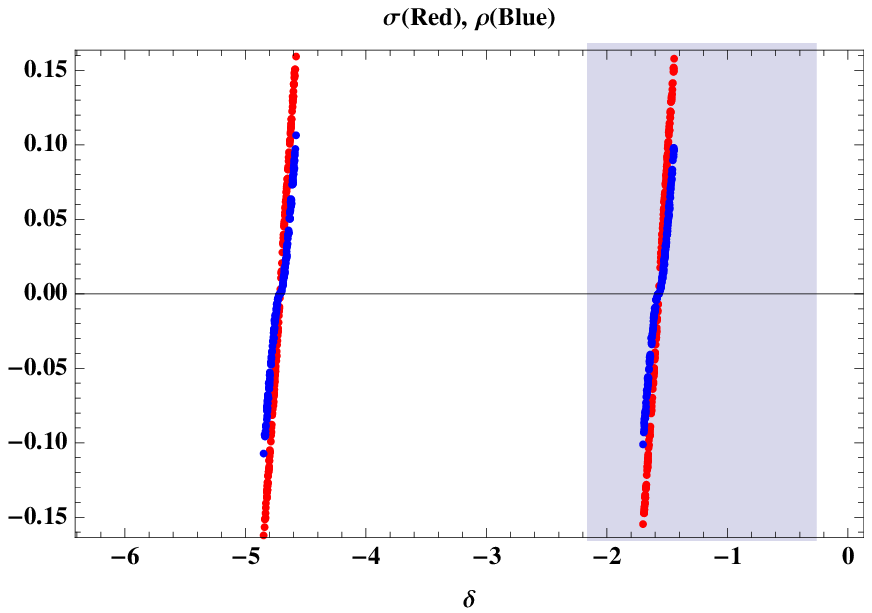} 
\includegraphics[width=80mm]{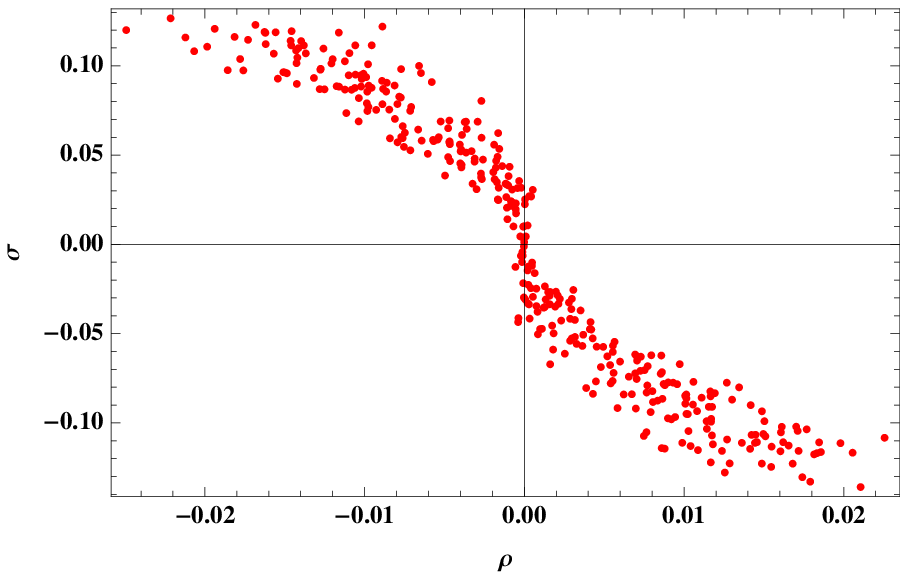} 
\includegraphics[width=80mm]{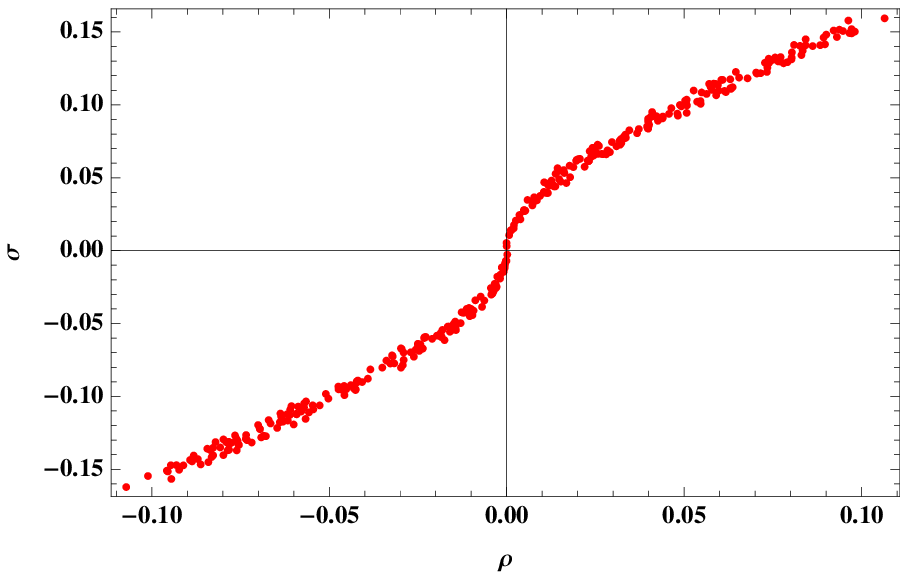} 
\includegraphics[width=80mm]{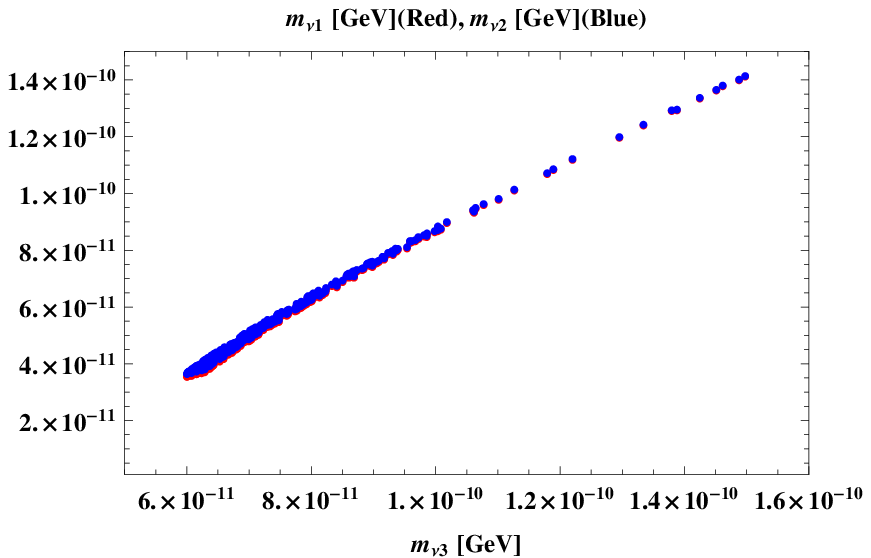}
\includegraphics[width=80mm]{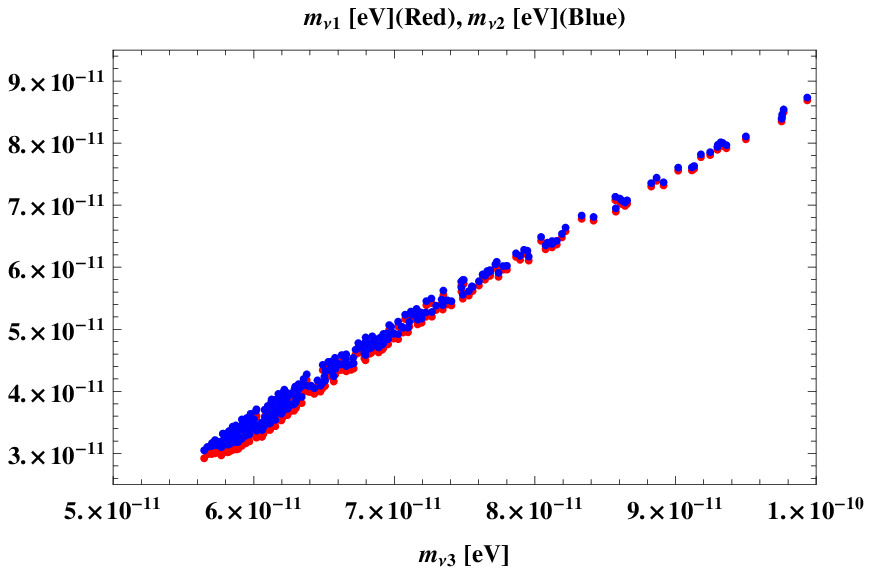}
\caption{The allowed regions in case of NH to satisfy the neutrino oscillation data, where the left-side of figures represent the case of type-II neutrino model, whereas the right-side ones do the case of canonical seesaw model.  
Upper figures represent the plots in terms of $\sigma-\delta$ with red color and $\rho-\delta$ with blue one,
and the bottom ones do  the plots in terms of $\rho-\sigma$,
the middle ones do the plots in terms of $m_{\nu_1}-m_{\nu_3}$ with red color and $m_{\nu_2}-m_{\nu_3}$ with blue one.
}  
\label{fig:NH-cases}
\end{center}\end{figure}

In Fig.~\ref{fig:NH-cases},
we represent the allowed regions in case of NH to satisfy the neutrino oscillation data, where the left-side of figures represent the case of type-II neutrino model, whereas the right-side ones do the case of canonical seesaw model.  
Upper figures represent the plots in terms of $\sigma-\delta$ with red color and $\rho-\delta$ with blue one,
where both of the cases have two solutions; $\delta=\pi/2,3\pi/2$, the latter solution of which is in favor of the current experimental data that is colored by  gray region, and the allowed region of $\rho$ is obtained as $|\rho|\lesssim 0.02(0.1)$ for Type-II(canonical) seesaw case.
On the other hand, the allowed region of $\sigma$ has similar behavior for the type-II case and canonical seesaw case; the former is within $|\sigma|\lesssim 0.12$ while the latter is within $|\sigma|\lesssim 0.15$.
The correlations between $\sigma$ and $\rho$ are shown with clearer way in the middle figures.
Bottom figures represent the plots in terms of  $m_{\nu_1}-m_{\nu_3}$ with red color and $m_{\nu_2}-m_{\nu_3}$ with blue one.
One finds that $m_{\nu_1}$ and $m_{\nu_2}$ are almost degenerate and their allowed region is $3.7 \times10^{-11}{\rm GeV}
\le m_{\nu_{1,2}}\le 1.4 \times10^{-10}{\rm GeV}$ in the type-II case, while  $3 \times10^{-11}{\rm GeV}\le m_{\nu_{1,2}}\le 8.8 \times10^{-10}{\rm GeV}$ in the canonical seesaw case. 
On the other hand, each of $m_{\nu_3}$ is predicted by $6 \times10^{-11}
\le m_{\nu_{3}} \le 1.5 \times10^{-10}{\rm GeV}$ in the type-II case, and $5.5\times10^{-11}{\rm GeV}\le m_{\nu_{1,2}}\le 1.0 \times10^{-10}{\rm GeV}$ in the canonical seesaw case.

\begin{figure}[h!]
\begin{center}
\includegraphics[width=80mm]{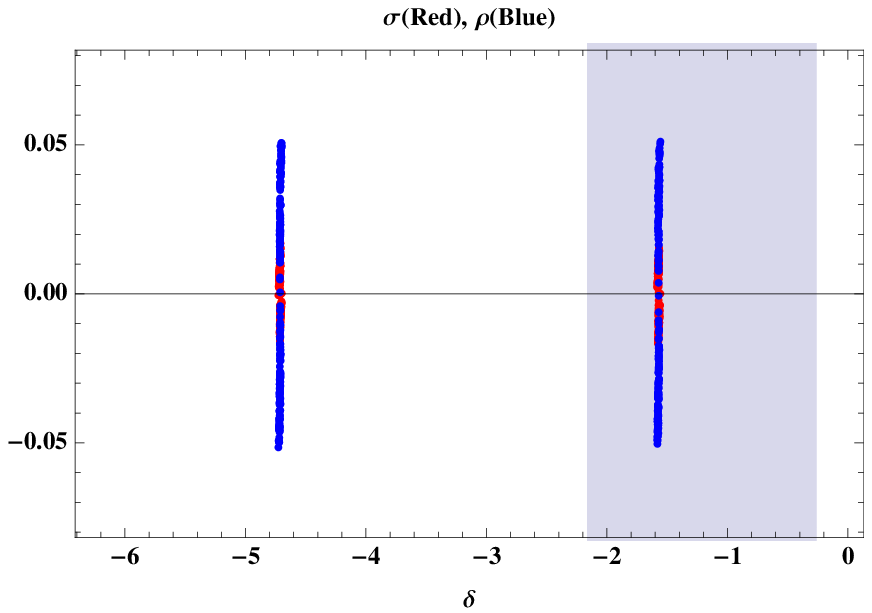} 
\includegraphics[width=80mm]{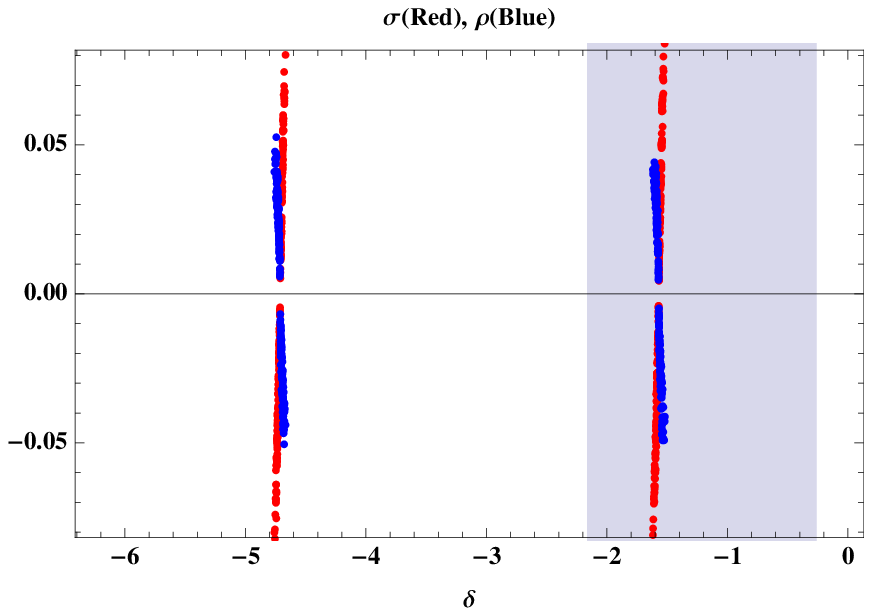} 
\includegraphics[width=80mm]{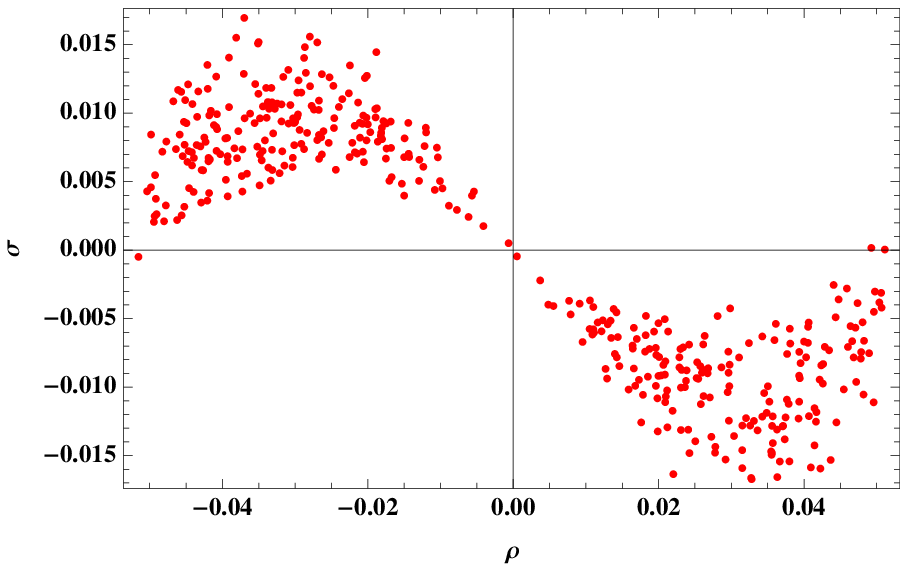} 
\includegraphics[width=80mm]{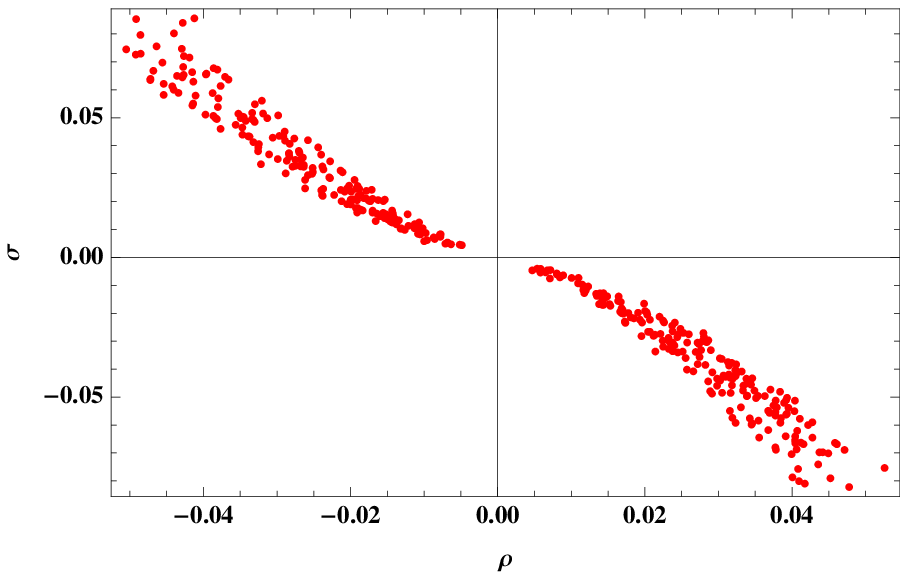} 
\includegraphics[width=80mm]{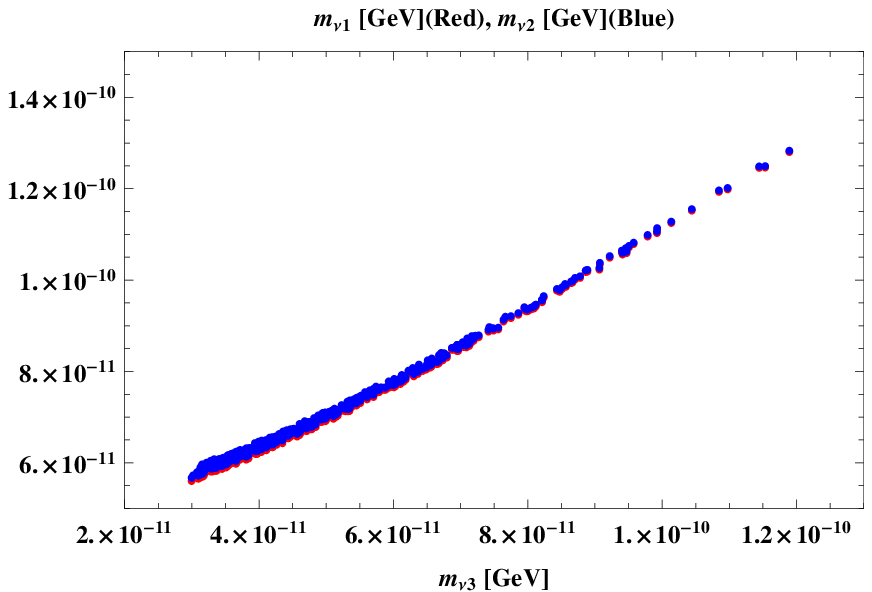}
\includegraphics[width=80mm]{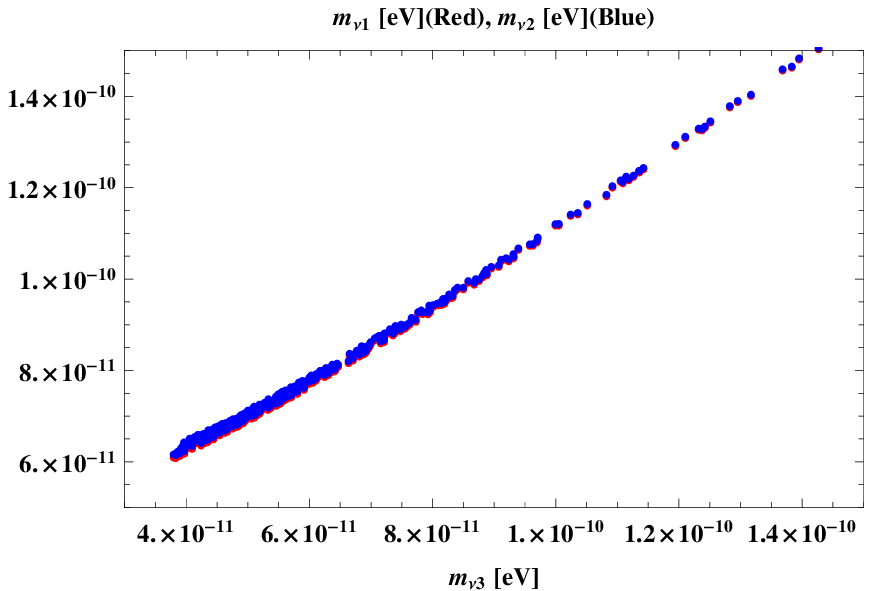}
\caption{The allowed regions in case of IH to satisfy the neutrino oscillation data, where the left-side of figures represent the case of type-II neutrino model, whereas the right-side ones do the case of canonical seesaw model.  
Upper figures represent the plots in terms of $\sigma-\delta$ with red color and $\rho-\delta$ with blue one,
and the bottom ones do  the plots in terms of $\rho-\sigma$,
the middle ones do the plots in terms of $m_{\nu_1}-m_{\nu_3}$ with red color and $m_{\nu_2}-m_{\nu_3}$ with blue one.
}  
\label{fig:IH-cases}
\end{center}\end{figure}

In figs.~\ref{fig:IH-cases},
we represent the allowed regions in case of IH to satisfy the neutrino oscillation data, where the left-side of figures represent the case of type-II neutrino model, whereas the right-side ones do the case of canonical seesaw model.  
Upper figures represent the plots in terms of $\sigma-\delta$ with red color and $\rho-\delta$ with blue one,
where both of the cases have two solutions; $\delta=\pi/2,3\pi/2$, the latter solution of which is in favor of the current experimental data that is colored by  gray region, and the common allowed region of $\rho$; $|\rho|\le0.05$.
On the other hand, the allowed region of $\sigma$ is different from  the type-II case and canonical seesaw case; the former is within $|\sigma|\le0.018$ while the latter is within $|\sigma|\le0.09$, which are shown with clearer way in the middle figures in~\ref{fig:IH-cases}.
Bottom figures represent the plots in terms of  $m_{\nu_1}-m_{\nu_3}$ with red color and $m_{\nu_2}-m_{\nu_3}$ with blue one.
One finds that $m_{\nu_1}$ and $m_{\nu_2}$ are almost degenerate and their allowed region is $5.5\times10^{-11}{\rm GeV}
\le m_{\nu_{1,2}}\le 1.3\times10^{-10}{\rm GeV}$ in the type-II case, while  $6\times10^{-11}{\rm GeV}\le m_{\nu_{1,2}}\le 1.5\times10^{-10}{\rm GeV}$ in the canonical seesaw case. 
On the other hand, each of $m_{\nu_3}$ is predicted by $3\times10^{-11}
\le m_{\nu_{3}} \le 1.2\times10^{-10}{\rm GeV}$ in the type-II case, and $3.5\times10^{-11}{\rm GeV}\le m_{\nu_{1,2}}\le 1.45\times10^{-10}{\rm GeV}$ in the canonical seesaw case. 

\subsection{Collider physics}

In this subsection, we discuss collider physics focusing on $Z'$ boson from $U(1)_{B-2 L_{\ell_2} - L_{\ell_1}}$ gauge symmetry.
Our $Z'$ can be produced at the LHC since it couples to quarks and it decays into SM fermions and/or singlet fermion pairs $\psi_i \psi_j$ if kinematically allowed.
The decay widths are obtained such that
\begin{align}
& \Gamma_{Z' \to \bar f f} = \frac{Q_{X_f}^2 g'^2}{12 \pi} m_{Z'} \left( 1- \frac{4 m_f^2}{m_{Z'}^2} \right)^{\frac12} \left(1 + \frac{2 m_f^2}{m_{Z'}^2} \right)  \\
& \Gamma_{Z' \to \psi_i \psi_j} = \frac{g'^2 (2V_{N_{i \ell_2}} V_{N_{\ell_2 j}}+ V_{N_{i \ell_1}} V_{N_{\ell_1j}})^2 }{16 \pi} m_{Z'} \lambda^{\frac12}(m_{Z'}; m_{\psi_i}, m_{\psi_j})  \nonumber \\
& \qquad \qquad \qquad \times \left( 1 - \frac{(m_{\psi_i}+ m_{\psi_j})^2}{m_{Z'^2}} - \frac{1}{3} \lambda^{\frac12}(m_{Z'}; m_{\psi_i}, m_{\psi_j}) \right), \\
& \lambda(m_1; m_2, m_3) \equiv 1 + \frac{m_2^4}{m_1^4} + \frac{m_3^4}{m_1^4} - \frac{2 m_2^2}{m_1^2} - \frac{2 m_3^2}{m_1^2} - \frac{2 m_2^2 m_3^2}{m_1^4},
\end{align}
where $f$ denote SM fermions and $Q_{X_f}$ is $U(1)_{B-2 L_{\ell_2} - L_{\ell_1}}$ charge of each fermions.
Then we show branching ratios (BRs) in Table.~\ref{tab:BRs} in which we ignored fermion masses assuming $Z'$ is much heavier than them.
Interestingly BRs for lepton modes depend on charge assignment in lepton sector where $BR(Z' \to \ell_2^+ \ell_2^-) = 4 BR (Z' \to \ell^+_1 \ell^-_1)$.
Since structure of neutrino mass matrix is also determined by the charge assignment we can test our scenario by comparing predictions in neutrino mass and 
pattern of BRs in leptonic decay mode of $Z'$.
\begin{table}[t!]
\begin{tabular}{|c||ccccc|}\hline\hline  
 & ~$q \bar q$~ & ~$\ell_2^+ \ell_2^-$~ & ~$\ell_1^+ \ell_1^-$~ & ~$\nu \bar \nu$~ & ~$\psi_i \psi_j$~ \\\hline\hline 
$BR$ & 0.028 & 0.33 & 0.083 & 0.21 & 0.21 \\ \hline 
\end{tabular}
\caption{ BRs for decay of $Z'$ where values for $\nu \bar \nu$ and $\psi_i \psi_j$ corresponds to sum of BRs for all possible modes, 
and we ignored fermion masses assuming $Z'$ is much heavier than them.}
\label{tab:BRs}
\end{table}

\begin{figure}[t]
\begin{center}
\includegraphics[width=70mm]{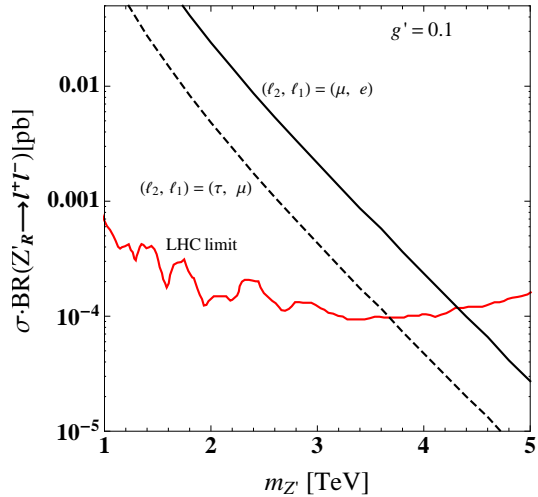} \qquad
\caption{The product of $Z'$ production cross section and $BR(Z'_R \to \ell^+ \ell^-)$ for $(\ell_2, \ell_1) = (\mu(\tau), e(\mu))$ with $g' = 0.1$ 
where region above red curve is excluded by the latest data~\cite{Aaboud:2017buh}. } 
  \label{fig:CX}
\end{center}\end{figure}

In Fig.~\ref{fig:CX}, we show $\sigma(pp \to Z') BR(Z' \to \ell^+ \ell^-)$ as a function of $m_{Z'}$ for the case of $(\ell_2, \ell_1) = (\mu, e)$ and $(\tau, \mu)$ 
where $BR(Z' \to \ell^+ \ell^-)$ correspond to sum of $e^+e^-$ and $\mu^+ \mu^-$ modes.
In addition, we compare the product of cross section and BR with the current limit from the data at the LHC experiments~\cite{Aaboud:2017buh}.
We find that $(\ell_2, \ell_1) = (\tau, \mu)$ case is less constrained and the $Z'$ should be heavier than several TeV when $g' \geq \mathcal{O}(0.1)$ in any cases.
On the other hand the collider constraint is the strongest for the case of $(\ell_2, \ell_1) = (\mu, e)$ or $(e, \mu)$.

\subsection{Dark matter candidate}
\begin{figure}[t]
\begin{center}
\includegraphics[width=70mm]{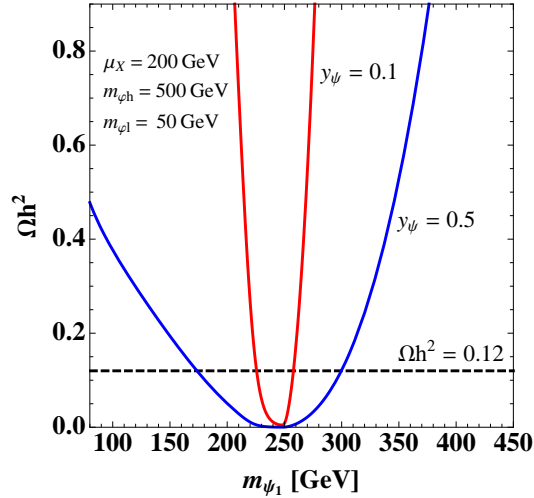} \qquad
\caption{Relic density of DM $\Omega h^2$ as a function of $m_{\psi_1}$ where some parameters are fixed as indicated in the plot. } 
  \label{fig:relic}
\end{center}\end{figure}

Here we discuss a  DM candidate in our model which depend on the hierarchy between $m_{0}$ and $D_{\psi}$ inside the loop diagram for neutrino mass generation. 
In case of $D_\psi<< m_0$, the lightest Majorana fermion is the DM candidate.
In case of $D_\psi\approx m_0$, either of fermion or boson  can be the DM candidate but we have to consider the coannihilation framework.
\footnote{In case of $D_\psi>> m_0$, the inert boson can be the DM candidate.
The allowed region to satisfy the relic density is at around $m_Z/2\lesssim M_X\lesssim m_W$~\cite{Cheung:2018itc} or $M_X\approx 500$ GeV~\cite{Hambye:2009pw}, where $M_X$ is the mass of DM.}
In our analysis, we focus on fermion DM case which is the lightest mass eigenstate $\psi_1$.  
The possible annihilation processes of DM are: (A) $\psi_1 \psi_1 \to Z' \to f \bar f$, (B) $\psi_1 \psi_1 \to h/\varphi_i \to f \bar f/W^+W^-/ZZ$, (C) $\psi_1 \psi_1 \to \varphi_i \varphi_i$, (D) $\psi_1 \psi_1 \to \varphi_h \to \varphi_l \varphi_l$.
The process (A) is annihilation via $Z'$ in s-channel. Since the gauge interaction with $Z'$ is constrained by collider experiments relic density can be explained for $2 m_{\psi_1} \simeq m_{Z'}$ due to resonance effect.
The process (B) is via Higgs or extra scalar portal case which also accommodates with observed relic density for $2 m_{\psi_i} \simeq m_{h,\varphi_i}$.
The third process (C) is annihilation process through Yukawa interaction among $\varphi_i$ and $\psi_1$.
The fourth process (D) is that heavier extra singlet scaler propagate in s-channel and it decays into pair of lighter singlet scalars. 
Here we discuss case (D) since it is specific and new process in the model due to the existence of at least two singlet scalars in realizing neutrino mass matrix.
In this analysis, we simply write relevant interactions schematically such that
\begin{equation}
\mathcal{L}_{\rm int} = y_{\psi_1} \varphi_h \bar  \psi_1^c P_R \psi_1 + \mu_{X} \varphi_h \varphi_l \varphi_l,
\end{equation}
where couplings $y_{\psi_1}$ and $\mu_X$ can be derived by writing original Lagrangian in terms mass eigenstates.
Then the squared amplitude for the process $\psi_1 \psi_1 \to \varphi_l \varphi_l$ via $s$-channel is given by
\begin{align}
|\bar {\cal M} (\psi_1 \psi_1 \to \varphi_l \varphi_l)|^2 & = (s- 4 m_{\psi_1}^2)
\left|\frac{ y_\psi \mu_X} {s-m_{\varphi_h}^2 + i m_{\varphi_h} \Gamma_{\varphi_h} }\right|^2  , 
\end{align}
where $\Gamma_{\varphi_h}$ is decay width of $\varphi_h$. 
The total and partial decay widths are given by
\begin{align}
& \Gamma_{\varphi_h} = \Gamma_{\varphi_h \to \psi_1 \psi_1} + \Gamma_{\varphi_h \to \varphi_{l} \varphi_{l}}, \\
& \Gamma_{\varphi_h \to \varphi_l \varphi_l} = \frac{\mu^2_X}{4 \pi m_{\varphi_h}} \sqrt{1 - \frac{4 m_{\varphi_l}^2}{m_{\varphi_h}^2}}, \\
& \Gamma_{\varphi_h \to \psi_1 \psi_1} = \frac{y_\psi^2}{8 \pi } m_{\varphi_h} \left(1 - \frac{4 m_{\psi_1}^2}{m_{\varphi_h}^2} \right)^{\frac32},
\end{align}
where kinematic condition is implicitly imposed and we simply assume $m_{\varphi_h} < 2 m_{\psi_{2,3}}$.
The relic density of DM can be estimated by~\cite{Edsjo:1997bg}
\begin{align}
&\Omega h^2
\approx 
\frac{1.07\times10^9}{\sqrt{g_*(x_f)}M_{Pl} J(x_f)[{\rm GeV}]},
\label{eq:relic-deff}
\end{align}
where $g^*(x_f\approx25)$ is the degrees of freedom for relativistic particles at temperature $T_f = M_X/x_f$ which is taken to be $g^* \simeq 100$, $M_{Pl}\approx 1.22\times 10^{19}$ GeV,
and $J(x_f) (\equiv \int_{x_f}^\infty dx \frac{\langle \sigma v_{\rm rel}\rangle}{x^2})$ is 
\begin{align}
J(x_f)&=\int_{x_f}^\infty dx\left[ \frac{\int_{4M_X^2}^\infty ds\sqrt{s-4 M_X^2} W(s) K_1\left(\frac{\sqrt{s}}{M_X} x\right)}{16  M_X^5 x [K_2(x)]^2}\right],\\ 
W(s)
&\equiv \frac{1}{16\pi}  \int_0^\pi \sin\theta
|\bar {\cal M}(\psi_1 \psi_1 \to \varphi_l \varphi_l) |^2 =
\frac{(s- 4 m_{\psi_1}^2)}{8 \pi} 
\left|\frac{ y_\psi \mu_X} {s-m_{\varphi_h}^2 + i m_{\varphi_h} \Gamma_{\varphi_h} }\right|^2 .
\label{eq:relic-deff}
\end{align}
In Fig.~\ref{fig:relic}, we show $\Omega h^2$ as a function of $m_{\psi_1}$ for $y_\psi = 0.1 (0.5)$ fixing some parameters as $m_{\varphi_h} = 500$ GeV, $m_{\varphi_l} = 50$ GeV and $\mu_X = 200$ GeV,
which is compared with $\Omega h^2 = 0.12$ taking from recent data of Planck observation $\Omega h^2 = 0.1187 \pm 0.0012$~\cite{Ade:2015xua}.
We then find that the observed relic density can be easily achieved with natural size of our parameters.
Note also that this result does not depend on $U(1)_{B - 2 L_{\ell_2} - L_{\ell_1}}$ charge assignment for leptons and can be applied in general case.

\section{Conclusions}
 We have studied predictive neutrino models with flavor dependent gauged  $U(1)_{B-2L_{\ell_2}-L_{\ell_1}}$ symmetry in which
we have successfully realized two-zero textures of neutrino mass matrix, depending on charge assignment for leptons and  the singlet scalar fields.
Then neutrino mass matrix is formulated which is generated at one-loop level, and lepton flavor violations are also considered induced by the Yukawa interactions for neutrino mass generation.
 We have also discussed some predictions for neutrino masses and CP-violating phases in a specific structure given by $(\ell_2, \ell_1) = (\mu, \tau)$ or $(\mu, e)$ and shown in Figs.~\ref{fig:NH-cases}, \ref{fig:IH-cases}.

 Then we have discussed phenomenologies such as collider physics and dark matter candidate in the model.
For collider physics, we have considered $Z'$ boson production at the LHC where branching ratio of $Z'$ decay mode depends on choice of charge assignment for leptons.
We have found that collider constraint in $(\ell_2, \ell_1) = (\tau, \mu)$ case is weaker than the other charge assignment. 
For dark matter, we have considered the case of fermionic dark matter which annihilate into new scalar boson from SM singlet scalar field.
We have shown that the observed relic density can be easily accommodated without strong constraints.

\section*{Acknowledgments}
\vspace{0.5cm}
Authors would like to thank Dr. Peiwen Wu for fruitful discussions.
H. O. is sincerely grateful for the KIAS member and all around.

\end{document}